# Optimization of Hybrid PV/Wind Power System for Remote Telecom Station

Subodh Paudel, J.N. Shrestha, Fernando J. Neto, Jorge A.F. Ferreira, Muna Adhikari

Graduate Student, Center for Energy Studies, Institute of Engineering, Tribhuvan University (TU), Nepal
Professor, Center for Energy Studies, Institute of Engineering, Tribhuvan University (TU), Nepal
Professor, Department of Mechanical Engineering, University of Aveiro (UA), Portugal
Junior Researcher, Innovative Energy and Environment Nepal (IEEN), Nepal

subodh.paudel@ntc.net.np, jnshrestha@gmail.com, fneto@ua.pt, muna.adhikari@ieen.org.np

**Abstract**

The rapid depletion of fossil fuel resources and environmental concerns has given awareness on generation of renewable energy resources. Among the various renewable resources, hybrid solar and wind energy seems to be promising solutions to provide reliable power supply with improved system efficiency and reduced storage requirements for stand-alone applications. This paper presents a feasibility assessment and optimum size of photovoltaic (PV) array, wind turbine and battery bank for a standalone hybrid Solar/Wind Power system (HSWPS) at remote telecom station of Nepal at Latitude ($27^023'50''$) and Longitude ($86^044'23''$) consisting a telecommunication load of Very Small Aperture Terminal (VSAT), Repeater station and Code Division Multiple Access Base Transceiver Station (CDMA 2C10 BTS). In any RES based system, the feasibility assessment is considered as the first step analysis. In this work, feasibility analysis is carried through hybrid optimization model for electric renewables (HOMER) and mathematical models were implemented in the MATLAB environment to perform the optimal configuration for a given load and a desired loss of power supply probability (LPSP) from a set of systems components with the lowest value of cost function defined in terms of reliability and levelized unit electricity cost (LUCE). The simulation results for the existing and the proposed models are compared. The simulation results shows that existing architecture consisting of 6.12 kW KC85T photovoltaic modules, 1kW H3.1 wind turbine and 1600 Ah GFM-800 battery bank have a 36.6% of unmet load during a year. On the other hand, the proposed system includes 1kW *2 H3.1 Wind turbine, 8.05 kW TSM-175DA01 photovoltaic modules and 1125 Ah T-105 battery bank with system reliability of 99.99% with a significant cost reduction as well as reliable energy production.

*Keywords: Hybrid PV/Wind energy system, Feasibility Study, Modeling, Optimization*

## INTRODUCTION

Telecommunication Networks have changed the way people live, work and play. Since many people around the world are connected by telecommunication networks, the challenge to provide reliable and cost effective power solutions to these expanding networks is indispensable for telecom operators. In remote areas, grid electricity is not available or is available in limited quantities. In the past, diesel generators with backup battery were used for powering these sites. These systems, usually located in areas with difficult accessibilities require regular maintenance and are characterized by their high fuel consumption and high transportation cost. Also, due to the rapid depletion of fossil fuel reserves and increasing demand of clean energy technologies to reduce the greenhouse gas emission ($CO_2$, $NO_X$, and $SO_X$) urgent search for alternative solutions for powering these sites is needed. Thus, stand alone renewable sources can be a feasible solution for powering these sites.

The various Renewable Energy Sources (RES) such as solar energy, wind energy, fuel cells, biodiesel and so on are used for telecommunications applications in the developing countries [1-4]. However, solar and wind are available freely and thus appears to be a promising technology to provide reliable power supply in the remote areas of Nepal. The intermittent nature of the solar and wind energy under varying climatic conditions requires a feasibility assessment and optimal sizing of hybrid solar and wind energy system. Without proper technical and financial feasibility study, the hybrid alternative energy systems previously installed in the remote areas showed a poor efficient design. The intent behind this paper is to design, optimize and analyze an effective hybrid PV-wind power system for a remote telecom station and to compare the existing system with the proposed new model.

The simple block diagram of the hybrid system is given below in figure (1). The hybrid solar and wind power system (HSWPS) works in two modes as: direct and indirect mode. The existing system implemented in Nepal Telecom (NT) at Dadakharka site consisting Code Division Multiple Access Base Transceiver Station (CDMA BTS), Very Small Aperture Terminal (VSAT) and Repeater Station and works in indirect mode, which means that power generation from renewable energy is stored in the battery and the energy stored in the battery is delivered to the load.

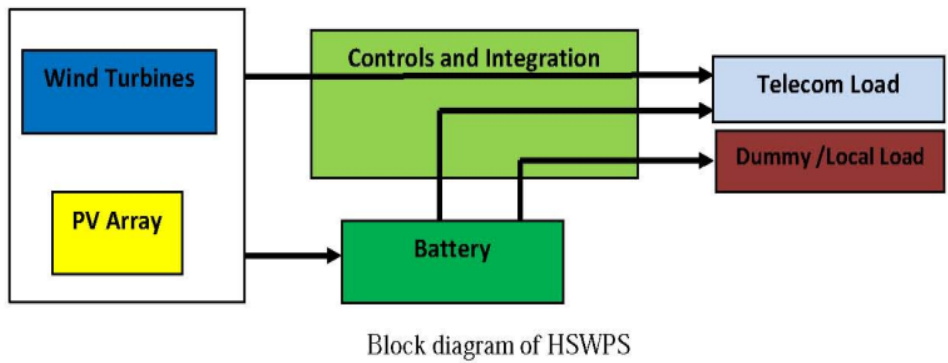

Fig. 1. Block Diagram of Hybrid Solar Wind Power System (HSWPS)

However, the proposed hybrid system working in direct mode is designed in such a way that if the power generation from wind and solar energy is higher than the load, the surplus power is stored in the battery. On the contrary, when the power generated from these resources is less than the power demanded by the load, the unmet power will be supplied through the battery storage system. In case of battery reaches to the maximum state of charge condition, the surplus energy is dissipated in the dummy load or local load in which the local people residing in that station lives.

Some authors [5-6] studied the feasibility assessment of photovoltaic and wind energy through RETScreen modeling software. Other authors [3, 7-10] carried the feasibility study of renewable energy sources with HOMER modeling software. From the literature [5-6], it is well known that RETScreen does not support and is not appropriate for hybrid system consisting of more than one renewable energy technology (e.g. PV and wind energy) although it is dedicated to feasibility analysis. In addition, analysis using Hybrid2 [11] emphasized system design with little focus on the pre-feasibility analysis of RES. HOMER gives more detailed information than the statistical models such as RETScreen and provides the optimization and sensitivity analysis with limited input [3, 7-10]. Moreover, HOMER is widely used for most of the RES based systems. Thus, based on the literature reviews, HOMER software is taken for the purposes of this study to carry the feasibility assessment.

**MATHEMATICAL SYSTEM MODELING**

A. PV Array Model

The PV generator is a non-linear device and is described by the I-V characteristics and by the equivalent circuit. There are many mathematical models developed to describe the behavior of PV [12-16]. In the present work, the "Four

Parameter Model", as: PV panel short circuit current, PV panel current at the maximum power point, PV panel maximum voltage at the maximum power point and PV panel open circuit voltage, which is widely used by the large software, is utilized [16]. The effect of change in temperature of PV cell [16] such as change in current due to the operating temperature and change in voltage due to operating temperature and PV panel operating cell temperature [17] based on energy balance is taken into consideration.

B. Wind Turbine Model

There are several factors that determine the output power from the wind turbine: the power output curve as a function of aerodynamic power efficiency, mechanical transmission and electrical energy conversion efficiency, wind speed distribution of the selected sites and the hub height of the wind tower. There are several existing models to estimate the wind turbine power such as- linear [18], cubic [19], quadratic [20-21], Weibull parameters [12] and so on. For the purposes of this study, Chou and Corotis model [21] is used since it is simple and widely used and does not required more specific information about the power characteristics curve and Weibull shape parameters. This model approximates the output power generated from the wind turbine for any wind speed distribution which takes account into cut-in wind speed, rated wind speed, cut-out wind speed and rated power of wind turbine.

The installation height of wind turbine has a large effect on the energy available from the system, so, a height adjustment is necessary for wind resources and is estimated based on power law [22].

The wind power available at the site for each hour is estimated assuming the wind speed as a function of Rayleigh Distribution [23]. Thus, the total average power available from the wind generator is calculated [22-23].

C. Temperature Model

Since the effect of temperature affects the output power of PV cells in very small amount, the temperature modeling plays a minor role in the design of HSWPS. There are various literature reviews on predicting hourly generation of temperature data from the collection of large number of data from weather stations and uses several methodology as Stochastic and Diurnal Model [24-25], Artificial Neural Network [26] etc.. Degelman Larry [25] used a Monte Carlo method to generate hourly weather data which includes both deterministic and stochastic models and takes all the parameters into consideration and is suitable for the purposes of this study to generate synthetic temperature profiles from the monthly average temperature of a year.

D. Battery Model

For any hybrid renewable energy systems, there is excess and deficit of energy at any instant of time. If the renewable resources system produces excess energy than the power demanded by the load, the extra energy should be stored. Also, if the renewable resources system produces less energy than the power demanded by the load, then storage system should satisfy the load. So, storage system is indispensable for any renewable energy stand-alone systems. There are various storage technologies to store the energy from renewable sources such as batteries [27-28], hydrogen combined with fuel cells [29-32]. Belfkira, Zhang and Barakat [28] which considers charging and discharging condition of battery with relevant technical specification is considered for the purposes of this study.

E. Reliability Model

The optimal sizing of hybrid system which meets the load demand is evaluated based on the power system reliability and system life cycle cost. The optimal solution of hybrid system can be best compromise with power reliability and system cost. The higher the power reliability, the higher will be the system cost and vice versa. There are various methods to calculate the reliability of the hybrid systems such as loss of load probability [33-34], loss of power supply probability [35], energy reliability index [36], least square method [37] and so on.

Since both solar and wind energy is intermittent in nature under the varying atmospheric conditions and the telecommunication load for the purposes of this study is constant for two different intervals i.e. half load and full load, the reliable power supply design is best suitable analysis for the purposes of this study. Considering this factor, loss of power supply probability (LPSP) [35] model is used for the purposes of this study and can be defined as the long-term average fraction of the total load that is not supplied by a stand-alone system.

A LPSP of 0 means that the load will be always be satisfied, and the LPSP of 1 means that the load will be never be satisfied [35]. Also, it is the probability that an insufficient power supply results when the hybrid PV and wind energy system is not able to satisfy the load demand. However, at any instant of time, the battery state of charge may be full or drops below the minimum state of charge and thus, waste energy and energy deficit is calculated [35]. Thus, the technical reliability can be further analyzed in terms of renewable energy contribution and excess energy percentage and is best estimated [35].

F. System Cost Model

The economic analysis of hybrid energy system is essential to know whether the system is affordable to produce unit price of power and to find the optimal configuration based on minimum cost. There are several economic measurements to know the system to be feasible or not: Net Present Cost [38], Annualized Cost [27], Life Cycle Cost [39-40], Levelized Cost of Energy [41-42] and so on. Based on the literature reviews, levelized cost of energy [42] is the most suitable model for the economic analysis of hybrid solar wind energy system and is considered as the economic model for purposes of this study.

G. Optimization Model

With the increase in number of optimization variables, the number of simulations also increases exponentially which takes more time and effort for the computation. Thus, it is necessary to find the optimal system configuration quickly and accurately using feasible optimization techniques. There are various optimization techniques used by the different authors for the design of HSWPS such as graphical construction methods [12], linear programming [43], iterative approach [44], genetic algorithm [27] and so on. Whichever the optimization method, the objective of each designer is to find the best optimal value for a given configuration. For the purposes of this study, Iterative Approach is used as an optimization technique which is simple to model and the output result produced by it is similar to the new evolved optimization techniques as genetic algorithm. The flowchart of optimization process is shown in figure (2).

The decision variables to find the optimal sizing are number of PV modules, number of wind turbines and number of batteries. The synthetic hourly solar radiation, wind speed and temperature data are generated from monthly average values. The assumption used for the system configurations are subject to the following constraints as:

$$1 \leq NBat \leq Nbatmax \quad (1)$$
$$1 \leq Npv \leq Npvmax \quad (2)$$
$$1 \leq Nw \leq Nwmax \quad (3)$$
$$SOCmin \leq SOC(t) \leq SOCmax \quad (4)$$

Where, *NBat* is the total number of batteries, *Nbatmax* is the maximum number of batteries, *Npv* is the total number of PV panels, *Npvmax* is the maximum number of PV panels, *Nw* is the total number of wind turbines, *Nwmax* is the maximum number of wind turbines, *SOC* is the state of charge of the battery, *SOCmin* is the minimum state of charge of the battery and *SOCmax* is the maximum state of charge of the battery.

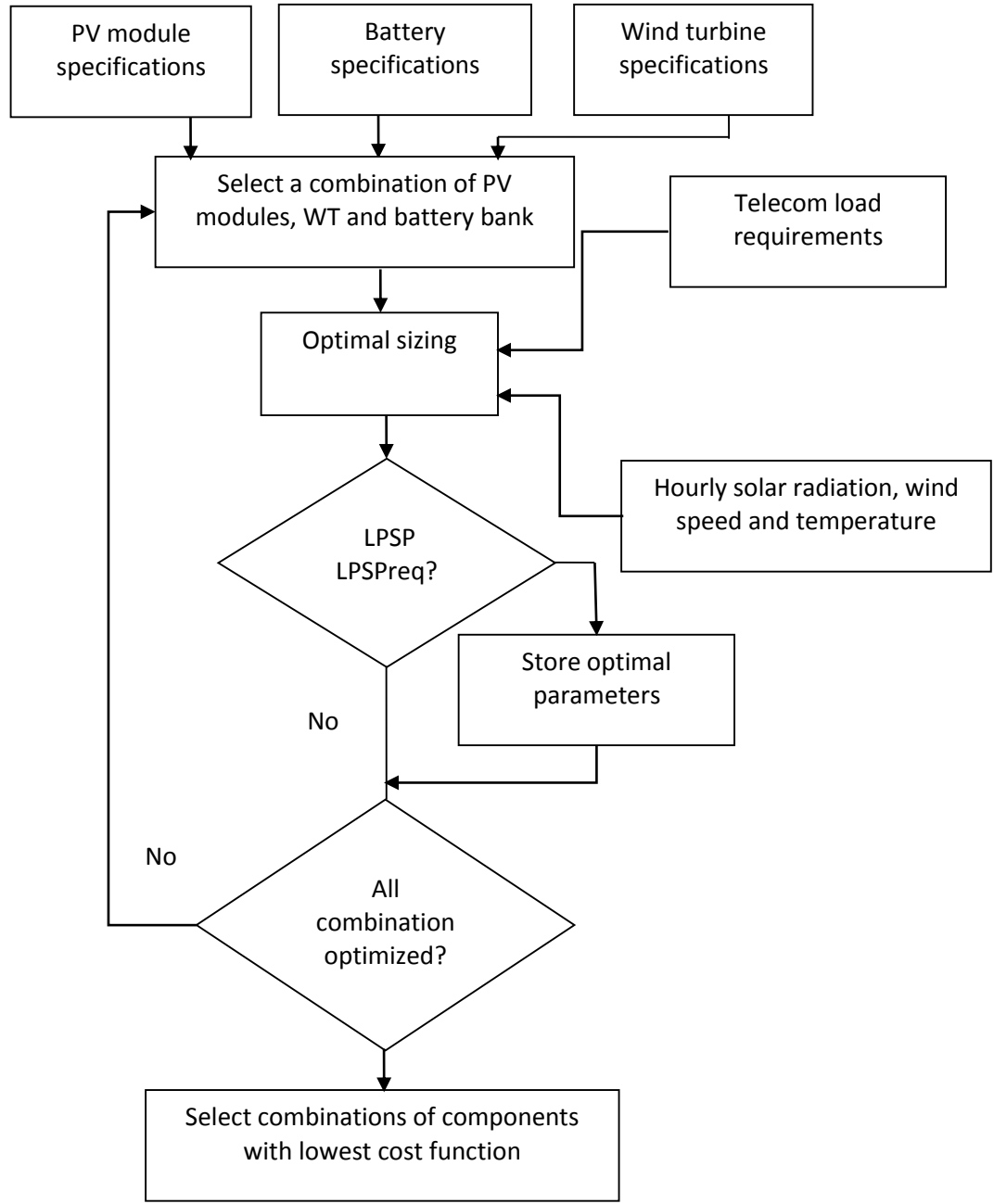

Fig. 2. Flowchart of proposed optimization methodology

The overall objective of the system is to achieve the minimum cost function (CF) which is defined in terms of reliability and levelized unit cost of electricity generation as shown in the equation (5).

$$CF = \text{Probability of reliability} * \text{Reliability} + \text{Probability of LUEC} * \text{LUEC} \qquad (5)$$

Reliability is calculated from LPSP and is given by equation (6). The configuration that meets the LPSP requirement is analyzed from objective function to find the optimal sizing configurations.

$$\text{Reliability} = 1\text{-LPSP} \qquad (6)$$

**RESULTS AND DISCUSSION**

The Dadakharka site located at Latitude (270 23' 50'') and Longitude (860 44' 23'') consisting 2C10 CDMA BTS, VSAT and Repeater Station of NT is taken as a case study for this purpose of this study. The existing system consists two 3.06 kW Kyocera KC85T PV array connected in parallel and the output of PV array is fed to charge regulator. Since, the bus voltage of telecommunication is 48V, for each array 4 panels are connected in series to provide fixed bus voltage and 18 panels are connected in parallel to meet the telecommunication load. The output of the charge regulator is connected to the battery so that it provides constant charging process. Instead of providing renewable power direct to load, all the energy generated from renewable resources is charged to the battery. The battery used is Narada GFM-800 of 800AH capacity with 2 battery bank where 24 batteries each of 2V are connected in series to meet bus voltage resulting in 48 batteries. Similarly, the power generated from the 1kW Hummer H3.1 DC wind turbine operating at 48V is fed to the hybrid solar/wind controller and the output of hybrid solar/wind controller is fed to the battery. Finally, the output voltage from the battery is fed to the charge controller so that it provides constant power to the load. Thus, the existing system with 6.12kW PV Array, 1 kW wind turbine and 1600AH battery bank provides the power supply to the remote station telecommunication load.

The daily telecom load profile is shown in figure (3). Figure (4) shows the optimization results of different configuration systems for varying wind speed and telecommunication load of existing system. From the figure (4) it is clear that only 6.12 kW KC85T PV system cannot meet the telecommunication load demand. The figure delineates that if the wind speed is below 4.5 m/s, only PV system is applicable to the telecom load upto 750Watt. Similarly, if the wind speed is above 7 m/s, only wind system is feasible for the all the load demand. However, most feasible configuration of the system is categorized by hybrid system consisting wind turbine and the PV array.

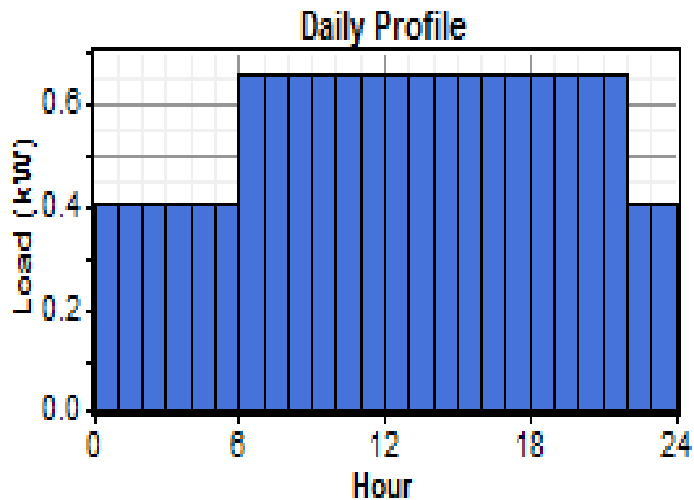

Fig. 3. Daily Telecom Load Profiles

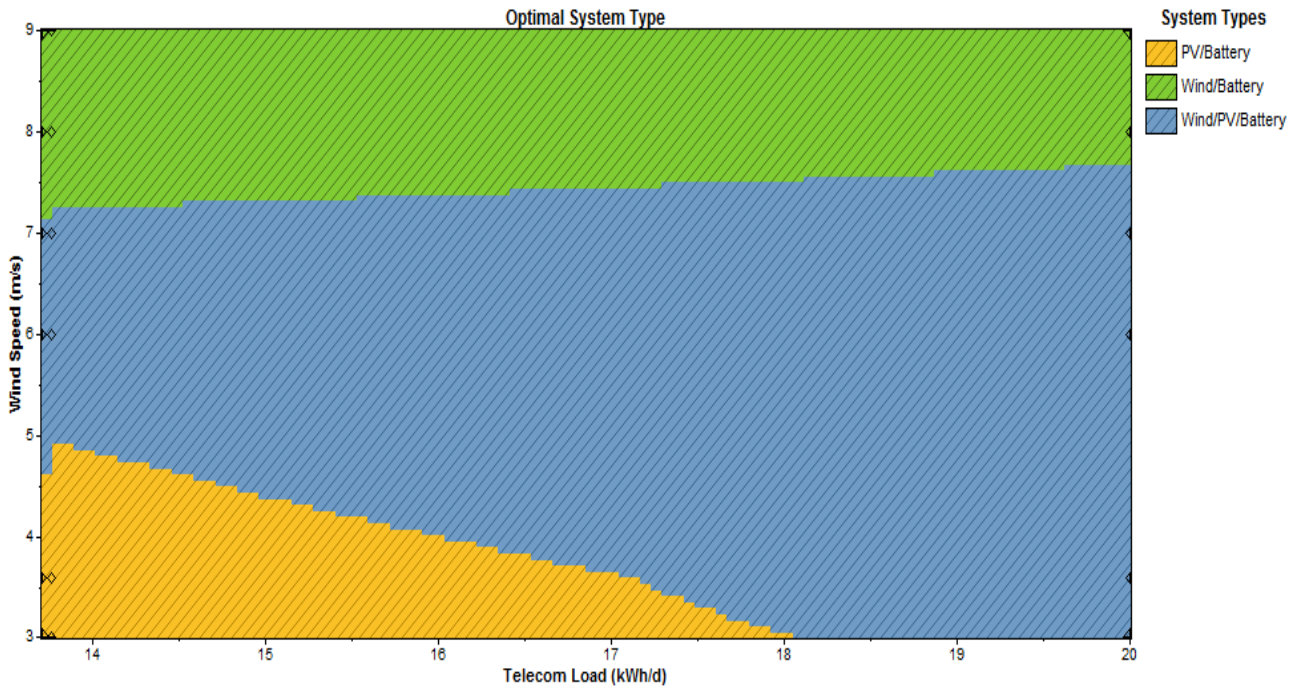

Fig. 4. Optimization results of different configuration systems

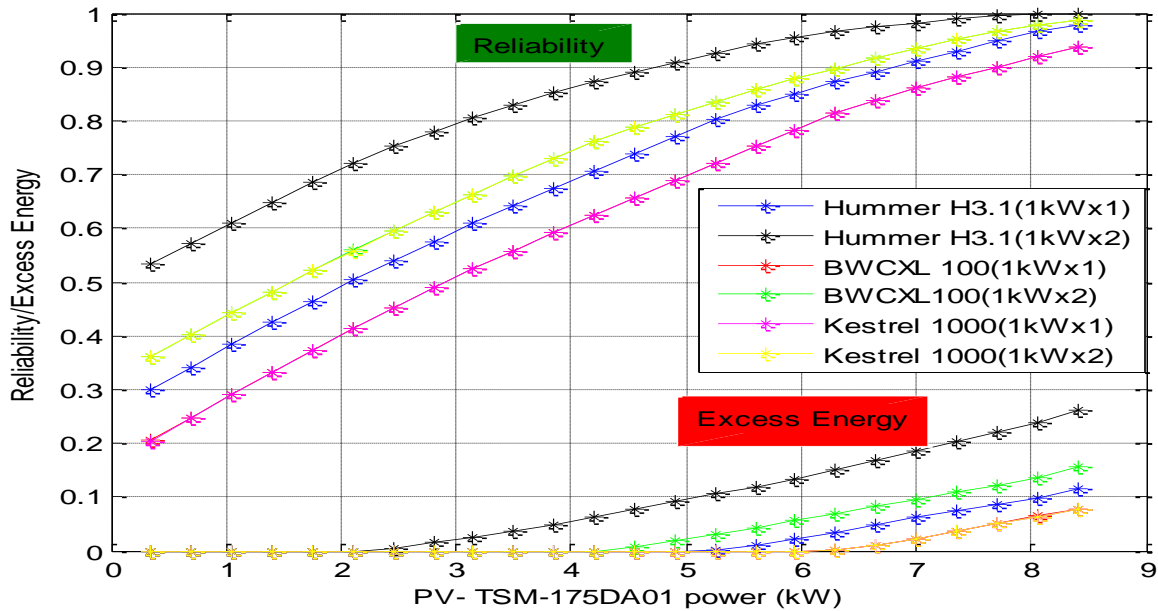

Fig. 5. Reliability and excess energy as a function of PV power

The simulation result shows that Hummer H3.1 (1kWx2) wind turbine, Trina Solar TSM-175DA01 (8.05kW) PV array and Trojan T-105 (1125Ah) justified the remote telecom load requirement of Dadakhara with reliability of 99.99% with a significant cost reduction as well as reliable energy production in the proposed system. On the contrary, the existing system containing Hummer H3.1 (1kW) wind turbine, Kyocera KC85T (6.12kW) PV array and Narada GFM-800 (1600Ah) have an unmet load of 36.6% during a year. The results can be analyzed in terms of excess energy, reliability and LUEC from different manufacturer specifications.

Figure (5) delineates the reliability and excess energy as a function of Trina Solar TSM-175DA01 power with Trojan T-195 (1125Ah) battery capacity for a different wind turbine. It can be observed from figure (3-3) that Hummer H3.1 has a better reliability than BWCXL 100 and Kestrel 1000 each of 1 kW capacity. From the figure it is clear that Kestrel 1000 (two 1kW capacity) has a less excess energy than rest of the system, however, the reliability of it is similar to the 1kW Hummer H3.1 wind turbine. The figure shows that Hummer H3.1 two 1kW wind turbine has a higher reliability than rest of the system with excess energy higher than rest.

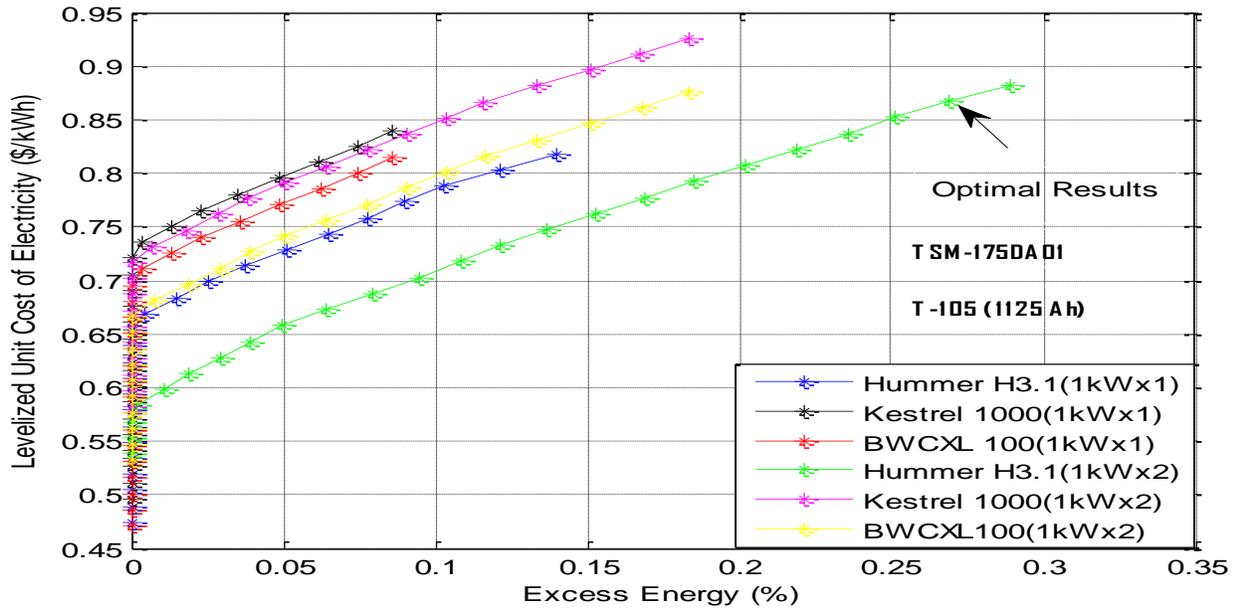

Fig. 6. LUEC as a function of excess energy for optimal configuration

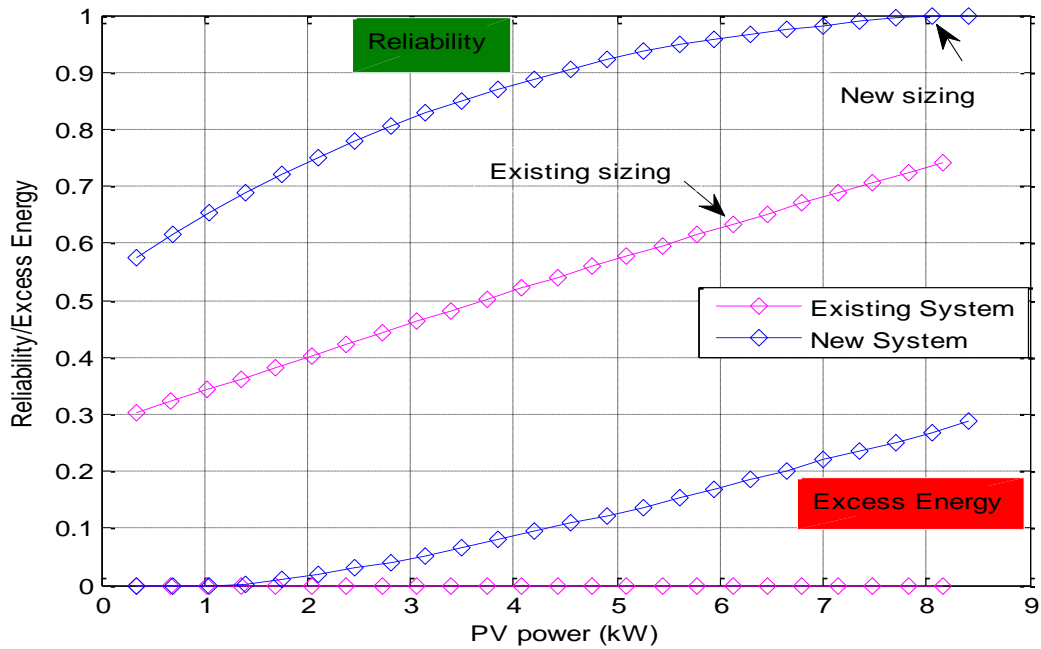

Fig. 7. Comparison of existing and new system

Figure (6) represent the LUEC as a function of excess energy for the optimal configuration system with TSM-175DA01 PV Array and battery T-105 of 1125Ah capacity. It can be observed from the figure that unit electricity cost generation from Hummer wind turbine is very cheap than rest of the turbines. The result shows that power generation from Kestrel is higher than rest of the system and the LUEC of optimal HSWPS configuration is found to be 0.86. Figure (7) represents the comparison of the existing NT system and the proposed system. It illustrates that existing system has deficient energy to fulfill the remote telecom load. The reliability is also poor than the newly proposed system which has reliability of 99.99 % with excess energy of 26.9% in a year. Figure (8) represents the hourly variation of state of charges of the battery T-105. The optimal configuration of the system consists of 40 batteries each of 225Ah with 5 battery bank in parallel resulting in a maximum capacity of 1125Ah in which remote telecom station load is always satisfied.

The simulation results shows that net present cost of existing and new system is $101,550 and $72,378. Similarly, The LUEC from the existing system is $1.22 and the proposed system is $0.88.

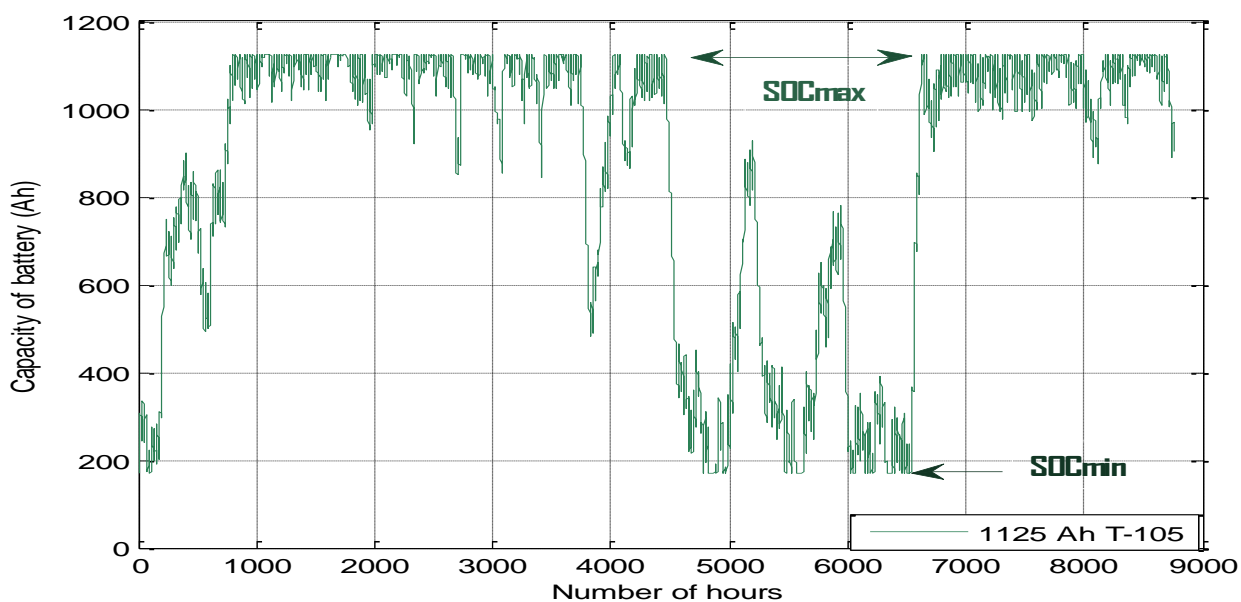

Fig. 8. Hourly variation of SOC of battery

**CONCLUSION**

Thus, it can be concluded that proposed system finds pragmatic applications in sizing and analyzing HSWPS. In addition, the size of the battery bank is reduced to 36% and overall 29% reduction in the total cost of the system. Moreover, the result shows the increment of reliability from 63.4% to 99.99% and decrement of LUEC from $1.22 to $0.88 in the proposed system and is thus the proposed system is economically and financially viable.

**REFERENCES**

1. Marian Vilsan, Irina Nita "A hybrid wind-photovoltaic power supply for a telecommunication system", *IEEE,* 1997.
2. Pragya Nema, Dr. Saroj Rangnekar, Dr. R.K. Nema, "Pre-feasibility study of PV-solar/wind hybrid energy system for GSM type mobile telephony base station in central India", *IEEE,* 2010.
3. Shafiqur Rehman, Luai M. Al-Hadhrami, "Study of a solar PV-diesel-battery hybrid power system for a remotely located population near Rafha, Saudi Arabia", *Energy*, vol. 35, pp. 4986-4995, 2010.

4. P. Bajpai, Prakshan N.P., N.K. Kishore, "Renewable Hybrid Stand-Alone Telecom Power System Modeling and Analysis", *IEEE, TENCON,* 2009.
5. Elizabeth Harder, Jacqueline MacDonald Gibson, "The costs and benefits of large-scale solar photovoltaic power production in Abu Dhabi, United Arab Emirates", *Renewable Energy*, vol. 36, pp. 789-796, 2011.
6. Y. Himri, A. Boudghene Stambouli, B. Draoui, "Prospects of wind farm development in Algeria", *Desalination*, vol. 239, pp. 130-138, 2009.
7. G.J. Dalton, D.A. Lockington, T.E. Baldock, "Feasibility analysis of stand-alone renewable energy supply options for a large hotel", *Renewable Energy*, vol. 33, pp. 1475-1490, 2008.
8. M.J.Khan, M.T. Iqbal, "Pre-feasibility study of stand-alone hybrid energy systems for applications in Newfoundland", *Renewable Energy*, vol. 30, pp. 835-854, 2005.
9. Ahmed M.A. Haidar, Priscilla N. John, Mohd Shawal, "Optimal configuration assessment of renewable energy in Malaysia", *Renewable Energy*, vol. 36, pp.881-888, 2011.
10. Getachew Bekele, Bjorn Palm, "Feasibility study for a standalone solar-wind-based hybrid energy system for application in Ethiopia", *Applied Energy*, vol.87, pp.487-495, 2010.
11. Andrew Mills, Said Al-Hallaj, "Simulation of hydrogen-based hybrid systems using Hybrid2", *International Journal of Hydrogen Energy*, vol. 29, pp. 991-999, 2004.
12. B.S. Borowy, Ziyad M. Salameh, "Methodology for Optimally Sizing the Combination of a Battery Bank and Pv Array in a Wind/PV Hybrid System", *IEEE Transactions on Energy Conversions*, vol.11, no.2, pp. 367-375, June 1996.
13. Markvard T., "Solar Electricity", *2$^{nd}$ ed. Wiley*, USA, 2000.
14. Zhou W., Yang H.X., Fang Z.H., "A Novel model for photovoltaic array performance prediction", *Applied Energy*, vol. 84, no. 12, pp. 1187-1198, 2007.
15. F. Lasnier, T.G. Ang, "Photovoltaic Engineering Handbook", Bristol, England, 1990.
16. M. Nikraz, H. Dehbonei, C.V. Nayar, "A DSP Controlled PV System with MPPT", *Australian Power Engineering Conference*, Christchurch, pp. 1-6, 2003.
17. John A. Duffie, William A. Beckman, "Solar Engineering of Thermal Process", *John Wiley & Sons, Inc,* 1991.
18. C.Bueno, J.A. Carta, "Technical-economic analysis of wind-powered pumped hydrostorage systems. Part I: model development", *Solar Energy*, vol.78, pp.382-395, 2005.
19. Faith O. Hocaoglu, Omer N. Gerek, Mehmet Kurban, "A novel hybrid (wind-photovoltaic) system sizing procedure", *Solar Energy*, vol.83, pp.2019-2028, 2009.
20. Lin Lu, Hongxing Yang and John Burnett, "Investigation on wind power potential on Hong Kong islands-an analysis of wind power and wind turbine characteristics", *Renewable Energy*, vol.27, pp.1-12, 2002.
21. Chou K.C., R.B. Corotis, "Simulation of Hourly Wind Speed and Array Wind Power", *Solar Energy*, vol. 26, pp. 199-212, 1981.
22. Rachid Belfkira, Cristian Nichita, Pascal Reghem, Georges Barakat, "Modeling and Optimal Sizing of Hybrid Renewable Energy System", *International Power Electronics and Motion Control Conference (EPE-PEMC), IEEE,* 2008.
23. Corotis RB, Sigl AB, Klein J, "Probability models for wind velocity magnitude and persistence", *Solar Energy*, vol.20, pp.483-493, 1978.
24. K.M. Night, S.A. Klein, J.A. Duffie, "A methodology for the syndissertation of hourly weather data", *Solar Energy*, vol.46, no.2, pp. 109-120, 1991.
25. Degelman Larry, "Simulation and uncertainty: weather predictions", *Advanced Bldg. Simulation, (Malkawi and Augenbroe, Ed.), Spon Press*, New York and London, chap.3, pp. 60-86, 2004.
26. Imran Tasadduq, shafiqur Rehman, Khaled Bubshait, "Application of neural networks for the prediction of hourly mean surface temperatures in Saudi Arabia", Renewable Energy, vol.25, pp. 545-554, 2002.
27. Wei Zhou, Hongxing Yang, Zhaohong Fang, "Battery behavior prediction and battery working state analysis of a hybrid solar-wind power generation system", Renewable Energy, vol.33, pp.1413-1423, 2008.
28. Rachid Belfkira, Lu Zhang, Georges Barakat, "Optimal sizing study of hybrid wind/PV/diesel power generation unit", Solar Energy, vol.85, pp.100-110, 2011.
29. B.D. Shakya, Lu Aye, P. Musgrave, "Technical feasibility and financial analysis of hybrid wind-photovoltaic system with hydrogen storage for Cooma", International Journal of Hydrogen Energy, vol.30, pp.9-20, 2005.
30. D.B. Nelson, M.H. Nehrir, C. Wang, "Unit sizing and cost analysis of stand-alone hybrid wind/PV/fuel cell power generation systems", Renewable Energy, vol.31, pp.1641-1656, 2006.
31. Kamaruzzaman Sopian, Mohd Zamri Ibrahim, Wan Ramli Wan Daud, Mohd Yusof Othman, Baharuddin Yatim, Nowshad Amin, "Performance of a PV-wind hybrid system for hydrogen production", Renewable Energy, vol.34, pp.1973-1978, 2009.
32. Vanhanen J.P., P.D. Lund, "Computational approaches for improving seasonal storage systems based on hydrogen technologies", International Journal of hydrogen energy, vol.20, pp. 575, 1995.
33. A.D. Bagul, Z.M. Salameh, B. Borowy, "Sizing of a stand-alone hybrid wind-photovoltaic system using a three-even probability density approximation", Solar Energy, vol.56, no.4, pp.323-335, 1996.
34. M.K. Deshmukh, S.S. Deshmukh, "Modeling of hybrid renewable energy systems", Renewable and Sustainable Energy Reviews, vol.12, pp.235-249, 2008.

35. S. Diaf, D. Diaf, M. Belhamel, M. Haddadi, A. Louche, "A methodology for optimal sizing of autonomous hybrid PV/wind system", Energy Policy, vol.35, pp.5708-5718, 2007.
36. Giuseppe Marco Tina, Salvina Gagliano, "Probabilistic modeling of hybrid solar/wind power system with solar tracking system", Renewable Energy, vol.36, pp.1719-1727, 2011.
37. S.Gomaa, A.K. Aboul Seoud, H.N. Kheiralla, "Design and Analysis of Photovoltaic and Wind Energy Hybrid Systems in Alexandria, Egypt", Renewable Energy, vol.6, no.5-6, pp.643-647, 1995.
38. Orhan Ekren, Banu Y. Ekren, Baris Ozerdem, "Break-even analysis and size optimization of a PV/wind hybrid energy conversion system with battery storage- A case study", Applied Energy, vol.86, pp.1043-1054, 2009.
39. HOMER Software Help Toolboxes, 2010.
40. G.C. Bakos, N.F. Tsagas, "Technoeconomic assessment of a hybrid solar/wind installation for electrical energy saving", Energy and Buildings, vol.35, pp.139-145, 2008.
41. J.P. Reichling, F.A. Kulachki, "Utility scale hybrid wind-solar thermal electrical generation: A case study of Minnesota", Energy, vol.33, pp.626-638, 2008.
42. A.Kaabeche, M. Belhamel, R. Ibtiouen, "Sizing optimization of grid-independent hybrid photovoltaic/wind power generation system", Energy, pp. 1-9, 2010.
43. R. Chedid, S. Rehman, "Unit sizing and control for hybrid wind-solar power systems", IEEE Transactions on Energy Conversion, vol.12, no.1. pp.79-85, 1997.
44. Hongxing Yang, Lin Lu, Wei Zhou, "A novel optimization sizing model for hybrid solar-wind energy power generation systems", Solar Energy, vol.81, no.1, pp.76-84, 2007.
45. Subodh Paudel, MSc Thesis, 2011, "Optimization of Hybrid PV/Wind Power System for Remote Telecom Station", *Institute of Engineering, Pulchowk Campus,* Nepal.